\documentclass[12pt,aps,prb,preprint]{revtex4-1}   % style for Physical Review B and AJP are similar

\usepackage{amsmath}    % need for subequations
\usepackage{graphicx}   % for figures

\begin{document}

\title{Feynman Integral and one/two slits electrons diffraction : an analytic study}
%Lines break automatically or can be forced with \\
\author{Mathieu Beau}
 \affiliation{School of Theoretical Physics, Dublin Institute for Advanced Studies, 10 Burlington Road,
Dublin 4}
 \email{mbeau@stp.dias.ie}  
\date{\today}

\begin{abstract}

In this article we present an analytic solution of the famous
problem of diffraction and interference of electrons through one
and two slits (for simplicity, only the one-dimensional case is
considered). 
In addition to exact formulas, 
we exhibit various approximations of the electron
distribution which facilitate the interpretation of the results.
Our derivation is based on the Feynman path integral formula and
this work could therefore also serve as an interesting pedagogical
introduction to Feynman's  formulation of quantum mechanics for
university students dealing with the foundations of quantum
mechanics.
\end{abstract}

\maketitle

\section{Introduction}

The quantum mechanical problem of diffraction and interference of
massive particles is discussed, though without detailed formulas,
by Feynman in his famous lecture notes.\cite{Feynman} A more
exact treatment, though still lacking in detail, is in his book
with Hibbs.\cite{FH} It was first observed experimentally by
J\"onsson in 1961.\cite{Jonsson} Moreover, there are also
experiments for neutrons diffraction for single and double slit 
(see \cite{Zeilinger} and the references therein)
and in quantum optics about interference between photons,
(e.g., \cite{Mandel}).  
The crucial point of this paper is to deal with different optical regimes: the usual
Fraunhofer regime, and (less commonly taught to students) the
Fresnel regime and intermediate regimes. 
Recall that these regimes depend on the distance between the slits and
the screen, where the Fraunhofer regime corresponds to the case
when the distance between the slits and the screen is infinite and
the other regimes appear when this distance is finite, the
intermediate and Fresnel regimes being distinguished by the value
of the Fresnel number $N_F\equiv 2a^2/\lambda L$, where $2a$ is the
width of the slit, $L$ is the distance between the screen and
the slit and $\lambda$ is the wave lenght of the electron. 
Thus, the purpose of this article is firstly to present the theory of the slit experiment
using the Feynman path integral formulation of quantum mechanics,
which may be of pedagogical interest as compared with the optical
Young experiment (Section II, III, IV), secondly to give an analytical
derivation of the final formulas for the intensity of the electron
on the screen, and to analyse these formulas using some
approximations based on the asymptotic behavior of the Fresnel
functions \cite{Abramowitz} occurring in these expressions (Section V). 
In particular we show how the physical parameters, especially the
Fresnel number, affect the form of the diffraction and
interference images. There exists some pedagogical papers 
about the multiple slit experiments 
(see e.g. \cite{Frabboni} for experimental and numerical discussions
and, \cite{Barut} for theoretical discussions using path integral formulation),
but surprisingly the approximations obtained in the Section V has never been published.

\section{Feynman formulation of quantum mechanics and why it could be of interest for students}

Students are often surprised to learn that under certain
physical conditions the experimental behavior of matter is
wave-like. In fact, this can present a didactical obstacle in
teaching the first course of quantum mechanics where the
principle of wave-particle duality can appear mysterious,
especially with electron diffraction experiments for one- and
two slits. Despite the interesting historical ramifications,
students often have many metaphysical questions which are not
answered satisfactorily in introductory quantum mechanics
courses. A formal and complete solution of the electron
diffraction problem based on Feynman's path integral approach
may be helpful in this regard. It could demystify the
diffraction experiment and clarify the principle of duality
between wave and particle by analogy with wave optics. In
addition, it could to be an interesting introduction to quantum
mechanics via the Feynman approach based on the notion of path
integral rather than the Schr\"odinger approach, highlighting
the analogy between quantum mechanics and optics.
Moreover, there is an interesting article that has recently been published 
in the \textit{European Journal of Physics Education}
that discusses using the Feynman path integral approach in secondary schools 
to teach the double slit experiment.\cite{Fanaro}

We begin by outlining Feynman's formulation of quantum mechanics.
We recall some of the fundamental equations before studying the
electron diffraction problem. For more detail and for historical
remarks about this theory, see Ref. \cite{FH},
\cite{Feynman2}, \cite{Feynman3}. We will give an equivalent formulation to that
based on Schr\"odinger's equation, but which is related to the
Lagrangian formulation of classical mechanics rather than the
Hamiltonian formulation. Consider a particle of mass $m$ under the
influence of an external potential $V[x(t),t]$ 
where $x(t)=(x^{(1)}(t),x^{(2)}(t),..,x^{(d)}(t))$ is the
(d-dimensional, $d\geq1$) coordinate of the particle at time $t$. The Lagrangian of
the particle has the simple form $$ L[x(t),\dot{x}(t),t] =
\frac{m}{2}\dot{x}(t)^2-V[x(t),t],
$$ where $\dot{x}(t) = dx(t)/dt$ denotes the velocity of the
particle at time $t$. If the particle is at position $x_i$ at time
$t_i$ and at $x_f$ at time $t_f>t_i$ then the \textbf{action} is
given by 
$$ S[x_f,t_f;x_i,t_i] \equiv
\int_{t_i}^{t_f} L[x(t),\dot{x}(t),t]\, dt. $$ 
In classical mechanics, the \textit{total variation} of the action $ \delta S $
for small variations of paths at each point of a trajectory
$\delta x(t) $ is zero, which leads to the Euler-Lagrange
equations well known to students of physics:
\begin{eqnarray*}
{d \over dt}{\partial L[x(t),\dot{x}(t),t]  \over
\partial\dot{x}^{(\alpha)}(t)} - {\partial L[x(t),\dot{x}(t),t]  \over
\partial x^{(\alpha)}(t)} = 0,\ \alpha=1,..,d\ .
\end{eqnarray*}

However, in quantum mechanics the Least Action Principle as
described above is generally not true. Thus, the concept of
classical trajectory is also no longer valid: the position as a
function of time is no longer determined in a precise way.
Instead, in quantum mechanics the dynamics only determines the
\textbf{probability} for a particle to arrive at a position $x_f$
at time $t_f$ knowing it was at a given position $x_i$ at time
$t_i<t_f$. In other words, knowing the \textit{state} of the
particle at time $t_i$, denoted by $|x_i,t_i\rangle$, the question
is: what is the probability of transition from the state
$|x_i,t_i\rangle $ to the final state $|x_f,t_f \rangle$? This
probability is given by the square of the modulus of the so-called
\textbf{amplitude} denoted $ \langle x_f,t_f|x_i,t_i \rangle$,
i.e. $|\langle x_f,t_f|x_i,t_i \rangle|^2$. The expression for the
amplitude in Feynman's formulation, though equivalent, differs
from Schr\"odinger's firstly because the Lagrangian formulation is
more general (there is not necessarily a Hamiltonian) and secondly
because it does not refer to waves. In Feynman's approach the
amplitude of transition is given by an \lq integral\rq\ over all
possible trajectories of a phase whose argument is the action
divided by Planck's constant $\hbar$, symbolically,
\begin{eqnarray*}
K(x_f,t_f;x_i,t_i) = \int Dx(t) e^{i S[x(t)]/\hbar},
\end{eqnarray*}
where $K(x_f,t_f;x_i,t_i) \equiv \langle x_f,t_f|x_i,t_i
\rangle$ is the amplitude, where the action functional is
$$ S[x(t)]\equiv\int_{t_i}^{t_f} L[x(t),\dot{x}(t),t]dt, $$ and where
$\int Dx(t)$ represent the \textit{path measure} to be interpreted
as follows: the integral formula is a short-hand for a limit of
multiple integrals:
\begin{eqnarray}\label{Klim}
K(x_f,t_f;x_i,t_i) &\equiv& \lim_{n\rightarrow\infty}
\frac{1}{(2i\pi \hbar\epsilon/m)^{d/2}}
\int_{R^d}\frac{dx_1}{(2i\pi \hbar\epsilon/m)^{d/2}}
\ldots\int_{R^d}\frac{dx_{n-1}}{(2i\pi \hbar\epsilon/m)^{d/2}}
{}\nonumber{}\\&& \quad \exp{\left(\frac{i}{\hbar}\epsilon
\sum_{k=1}^{n}(\frac{m(x_{k}-x_{k-1})^2}{2\epsilon^2}-V(x_k))
\right)},
\end{eqnarray}
where $\epsilon=(t_f-t_i)/n$, $x_0 = x_i$ et $x_n = x_f$.

\begin{figure}
 \centering
\includegraphics[width=130mm]{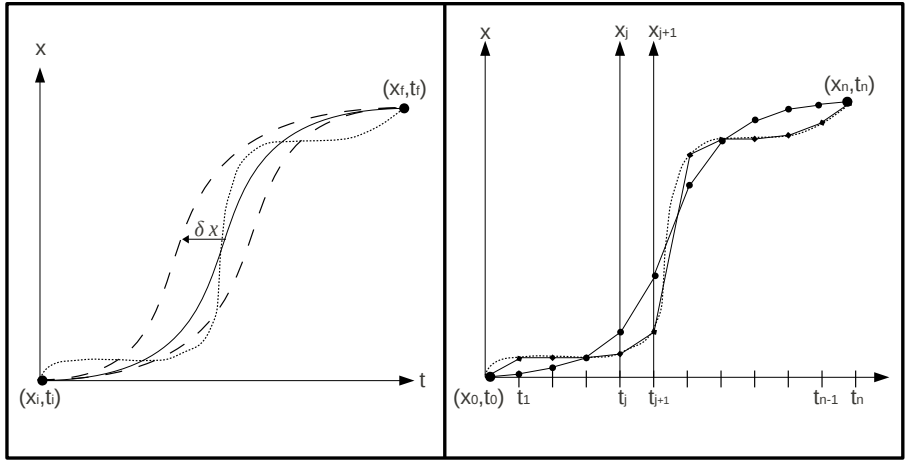}
\caption{\label{Fig.0} Representation of paths between $(x_i,t_i)$ and $(x_f,t_f)$, 
for the continuous case at the left (integral over continuum paths, see (\ref{Klim})) 
and for the discrete case at the right (integral over discrete paths before the continuum limit, see (\ref{Kn})).
In the left figure, the continuous line is the classical path, 
the dashed lines are two paths obtained by variations around the classical paths
and the dotted line is an arbitrary path.
In the picture at the right the discrete path for fixed $n$ is shown together 
with the classical discrete path.} 
\end{figure}

To understand this formula, we recall some ideas from Feynman's
thesis \cite{Feynman2}. Consider the particle at a point $x$ at
time $t$. Suppose that the particle changes position by an amount
$\delta x$ during an infinitesimal time interval $\delta t$. The
action for this time interval can be written as $S[x+\delta,t+\Delta t;x,t] 
\simeq L[x,\frac{\delta x}{\delta t},t] \delta t$.
We define the amplitude of transition between the states
$|x,t\rangle $ and $|x+\delta x,t+\delta t\rangle $ as $$ \langle
x+\delta x,t+\delta t|x,t\rangle \equiv
\frac{1}{(2i\pi \hbar\epsilon/m)^{d/2}}
e^{\frac{i}{\hbar}S[x+\delta x,t+\delta t,x,t]}
\simeq\frac{1}{(2i\pi \hbar\epsilon/m)^{d/2}}
e^{\frac{i}{\hbar}L[x,\frac{\delta x}{\delta t},t]\delta t} .
$$ Then, if we divide the time interval $[t_i, t_f]$ into a
sequence of small intervals $t_i=t_0 < t_1 < t_2 < \ldots <
t_{n-1} < t_f=t_n $, where $t_k=t_i + k(t_f-t_i)/n$, the
transition amplitudes are multiplied (because successive events
are independent): $$ \langle x_f,t_f|x_i,t_i\rangle = \langle
x_f,t_f|x_{n-1},t_{n-1}\rangle \ldots \langle
x_2,t_2|x_1,t_1\rangle  \langle x_1,t_1|x_i,t_i\rangle\ .$$
Integrating over all possible values of the intermediate positions
$x_k$ leads to the formula (\ref{Klim}), see Fig. \ref{Fig.0}. The intuition associated
with this formula is as follows: we integrate the phase over all
possible trajectories of the particle; unlike the classical case,
the particle can follow paths that differ from the classical
trajectory which minimizes the action. (Paths far away from the
classical path usually make negligible contributions.) This image,
though based on the classical image of a trajectory, illustrates
the change in the mathematical description of the particle
(wave-like behaviour) which can no longer be represented as a
material point as its trajectory is not clearly defined with
certainty.

The integral equation describing the evolution of the wave
function $\Psi(x,t)$ (i.e. the state $|x,t\rangle$) over time, is
given by
\begin{equation}\label{ShroInt}
\Psi(x',t')=\int_{R^d} K(x',t';x,t)\Psi(x,t) dx\ .
\end{equation}
It can be shown that the \lq wave function\rq\ $\Psi(x,t)$ is also
the solution of Schr\"odinger's equation:
$-\frac{\hbar^2}{2m}\Delta_x\Psi(x,t)+V(x,t)\Psi(x,t)=i\hbar\partial_t\Psi(x,t)$,
where $\Delta_x$ is the $d$-dimensional Laplacian and
$\partial_t\equiv\partial/\partial t$. For a proof of the
equivalence between both formulations, see
\cite{FH},\cite{Feynman2}.

A useful example is the free particle amplitude calculation in
one dimension. We simply replace the Lagrangian by the free
Lagrangian, i.e. without potential, and use equation
(\ref{Klim}) to get :
\begin{equation}\label{Kfree}
K_0(x',t';x,t)=\frac{1}{\sqrt{2i\pi \hbar(t'-t)/m}}
e^{i\frac{m(x'-x)^2}{2\hbar (t'-t)}}\ .
\end{equation}
Let us derive formula (\ref{Kfree}) starting from the general formula (\ref{Klim})
applied to a free particle in 1-dimension.
We divide the integral, for fixed $n$, we have to integrate the phase $e^{i S[x_0,..,x_n]/\hbar}$
over the intermediate positions $x_j,\ j=1,..,n-1$ at times $t_j=j\epsilon,\ \epsilon=(t_f-t_i)/n$ 
between $x_0=x_i$ and $x_n=x_f$, where $S[x_0,..,x_n]=\frac{m}{2}\sum_{j=1}^n\frac{(x_j-x_{j-1})^2}{\epsilon}$,
multiplied by the constant $\left(\prod_{j=1}^{n}\sqrt{m/2i\pi\hbar\epsilon}\right)$. 
We need to calculate, for a fixed number of subdivision $n$,
the multiple integral:
\begin{eqnarray}\label{Kn}
K_0^{(n)}&&(x_n,t_n;x_0,t_0)\equiv
{}\nonumber{}\\{}&&\frac{1}{\sqrt{2i\pi \hbar\epsilon/m}}\int_{-\infty}^{\infty}
\frac{dx_1}{\sqrt{2i\pi \hbar\epsilon/m}}
\ldots\int_{-\infty}^{\infty}\frac{dx_{n-1}}{\sqrt{2i\pi \hbar\epsilon/m}}
\exp{\left(i\sum_{j=1}^{n}(\frac{m(x_{j}-x_{j-1})^2}{2\hbar\epsilon})
\right)}\ .
\end{eqnarray}
We perform the integration one stage at a time (from $j=1$ to $j=n-1$),
using the well known formula for the convolution of two Gaussians.
Given two Gaussians $g_{\sigma_\nu}(x)\equiv\frac{1}{\sqrt{2\pi\sigma_\nu^2}}
e^{-\frac{x^2}{2\sigma_\nu^2}},\ \nu=1,2$, we have:
\begin{eqnarray}\label{GaussianConvo}
\int_{-\infty}^{\infty}g_{\sigma_1}(x-x')g_{\sigma_2}(x')=g_{\sqrt{\sigma_1^2+\sigma_2^2}}(x)\ . 
\end{eqnarray}
Hence, in (\ref{Kn}), we can calculate the integral with respect to $x_1$,
changing variable to $x'_1=x_1-x_0$, obtaining:
\begin{eqnarray*}
&&\frac{1}{\sqrt{2i\pi \hbar\epsilon/m}}\int_{-\infty}^{\infty}\frac{dx_1}{\sqrt{2i\pi \hbar\epsilon/m}}
e^{i\frac{m(x_{2}-x_{1})^2}{2\hbar\epsilon}}e^{i\frac{m(x_{1}-x_0)^2}{2\hbar\epsilon}}
{}\\{}&=&\frac{1}{\sqrt{2i\pi \hbar(2\epsilon)/m}}e^{i\frac{m(x_{2}-x_0)^2}{2\hbar(2\epsilon)}}\ .
\end{eqnarray*}
Proceeding by recursion, we get (\ref{Kn}) :
\begin{eqnarray*}
K_0^{(n)}(x_n,t_n;x_0,t_0)&=&\frac{1}{\sqrt{2i\pi \hbar(n\epsilon)/m}}e^{i\frac{m(x_{n}-x_0)^2}{2\hbar(n\epsilon)}}\ ,
\end{eqnarray*}
and hence (\ref{Kfree}) in the limit $n\rightarrow\infty$.

Clearly the 3-dimensional propagator is a product of 1-dimensional propagators:
\begin{eqnarray}\label{K3D}
K_{0}^{(3D)}(\overrightarrow{x}',t';\overrightarrow{x},t)&=&K_0(x',t';x,t)K_0(y',t';y,t)K_0(z',t';z,t)
{}\nonumber\\{}&=&\left(\frac{m}{2i\pi\hbar(t'-t)}\right)^{3/2}
e^{i\frac{m|\overrightarrow{x}'-\overrightarrow{x}|^2}{2\hbar(t'-t)}}\ , 
\end{eqnarray}
where $\overrightarrow{x}=(x,y,z)$ is the position vector in 3-dimensions.
Note that for reasons to be discussed below,
we shall not make use of the 3-dimensional propagator in the example 
of election diffraction for one and two slits. 
We remark for completeness, that the propagator (\ref{Kfree})
can be calculate by other methods, for example.\cite{Kleinert}

\section{Application to the Problem of Electron Diffraction and Interference}

Consider an electron source at $(x,y,z) = (0,0,0)$, and two slits
at $z=D$ of width $2a$ and centered respectively at $x=+b$ and
$x=-b$, c.f. Fig. \ref{Fig.1} . For the diffraction experiment with
a single slit, just replace the system with a slit of size $ 2a $
centered at $x=0$. At $ z = D + L $ is a screen on which electrons
are recorded (or some other recording device). For more details
about the experimental realization of the system, see Ref. \cite{Frabboni} 
Note that we neglect gravity for this
problem. In addition, we assume that in the direction orthogonal
to the plane in Fig. \ref{Fig.1} (the $x-z$-plane), the slot is long
enough to neglect diffraction effects; we consider only the
horizontal (in-plane) deflection of the beam in order not to
complicate the formulas. 

Then we can reduce the dimension of the propagator (\ref{K3D}) by integrating over $y$:
\begin{eqnarray}\label{K2D}
K_{0}^{(2D)}(\overrightarrow{r}',t';\overrightarrow{r},t)&=&
\int_{-\infty}^{\infty}dy\ K_{0}^{(3D)}(\overrightarrow{x}',t';\overrightarrow{x},t)
{}\nonumber\\{}&=&\frac{m}{2i\pi\hbar(t'-t)}
e^{i\frac{m|\overrightarrow{r}'-\overrightarrow{r}|^2}{2\hbar(t'-t)}}\ ,  
\end{eqnarray}
where $\overrightarrow{r}=(x,z)$ is the position vector in the two dimensional plane orthogonal to the $y$-axis.

One can suggest different models for the slits given by
distribution functions (e.g. Gaussian functions or \lq door\rq\
functions). We focus on the more realistic model of \lq door\rq\
functions (for the Gaussian function model, see Ref. \cite{FH}):
\begin{eqnarray}\label{Porte}
\chi_{[b-a,b+a]}(w)=\begin{cases}
0 & \text{$w > b+a,\ w < b-a$} \\
1 & \text{$b-a<w<b+a$}
\end{cases}
\end{eqnarray}
The question is: what is the probability of finding the electron
at the point $ x $ on the screen knowing that it started at the
point $ (x = 0, z = 0) $? More precisely, suppose that the
source emits electrons in large numbers, though small enough so
that the distance between electrons is such that interactions
can be neglected (no correlations). What is the intensity of
electrons on the screen as a function of position $x$? The two
questions are related since under these circumstances the motion
of the electrons is independent (without mutual interaction,
given that the density of the beam is very low). Indeed, the
intensity curve is obtained by simply multiplying the
probability curve for an electron by the number of electrons
emitted per unit time.
{As explained above formula (\ref{Klim}),
the probability is given by the square of the amplitude which 
we will compute using the propagator (\ref{K3D}).}\\

As explained in (\ref{K2D}), we now take $d=2$.
In fact, we now argue that we can reduce the dimension further. 
For the $z$-direction, we really onght to use the two-dimensional propagator
but we can reasonnably consider (as it is usually done\cite{Frabboni},\cite{FH}) 
that the propagator is a product of two independant propagator
in both directions $x$ and $z$, where the last one is equal to a constant.

To see that, let us discuss the conceptual experiment for one slit.\cite{FH} 
We consider that the problem is divided 
in two separate motions, one starting from the source to the slit during the time $T$
and the other one starting from the slit to the screen during the time $\tau$. 
Then, we would like to compute the probability amplitude for the electron
from the source, at the intial position $(x=0,z=0)$ at the time $t=0$, 
to go to the screen, at the final position $(x,z=D+L)$ at the time $T+\tau$, 
knowing that it goes through the slit, 
at the intermediate position $(w,z=D),\ b-a<w<b+a$, at the time $t=T$. 
In fact, by the laws of quantum mechanics, there is no reasons
to separate the motions in two independant parts, since we don't know
where is the particle at the time $T$.In other words we don't
know when the particle goes through the slit.
Nevertheless, we can consider that this classical image is appropriated
to study the problem. Indeed, the electron have in the $z$-direction 
the momentum $p_z=\hbar k_z$ ($k_z$ is the wave vector), 
which is related to the classical velocity $v_z=D/T=L/\tau$, 
where $D$ is supposed to be very large compared to the dimensions in the $x$-direction,
more precisely, $x,a,b\ll D,L$. In addition we suppose that the wave lenght $\lambda$, 
which is approximatively equal to the $z$-direction wave lenght $\lambda\simeq\lambda_z=2\pi\hbar/(mv_z)$, 
is small compared to the distances $\lambda\ll D,L$. 
Thus, the motion is approximatively classical in the $z$-direction 
and we can separate the problem as two independant motions.
Notice that, quantum-mechanically, it is possible for the particle to go through the slit 
severals time before strike the screen,\cite{Yabuki}
but that the probability of this event is relatively small.

Now, let us compute the amplitude of the transition for the particle 
starting at the point $(x=0,z=0)$ at the time $t=0$,
going through one slit at the position $(w,z=D),\ b-a<w<b+a$ at the time $t=T$
and arriving at the position $(x,z=L+D)$ at the time $t=T+\tau$: 
\begin{eqnarray*}
K_{a,b}^{(2D)}((x,L+D),T+\tau;&&(0,0),0)
\equiv\int_{-\infty}^{\infty}dw\ \chi_{[b-a,b+a]}(w)\times
{}\nonumber\\{}&&K_0^{(2D)}\left((x,L+D),T+\tau;(w,D),\tau\right)K_0^{(2D)}\left((w,D),\tau;(0,0),0\right)\ .
\end{eqnarray*}
Hence, the explicite formula is given by:
\begin{eqnarray}\label{K2DFormula}
K_{a,b}^{(2D)}((x,L+D),&&T+\tau;(0,0),0)=
{}\nonumber\\{}&&\frac{e^{i\frac{mD^2}{2\hbar T}}}{\sqrt{2i\pi\hbar T/m}} 
\frac{e^{i\frac{m L^2}{2\hbar\tau}}}{\sqrt{2i\pi\hbar\tau/m}}
\int_{b-a}^{b+a}dw\ \frac{e^{i\frac{m (x-w)^2}{2\hbar\tau}}}{\sqrt{2i\pi\hbar\tau/m}}
\frac{e^{i\frac{m w^2}{2\hbar T}}}{\sqrt{2i\pi\hbar T/m}}\ .
\end{eqnarray}
Consequently, the two-dimensional propagator is the product of two independant 
one-dimensional propagators in the $x$ and in the $z$-directions, 
and since the propagator in $z$-direction is a constante 
(see the right hand sides of (\ref{K2DFormula})):
\begin{eqnarray}\label{Kz}
K_z(L+D,T+\tau;0,0)=\frac{e^{i\frac{mD^2}{2\hbar T}}}{\sqrt{2i\pi\hbar T/m}}
\frac{e^{i\frac{m L^2}{2\hbar\tau}}}{\sqrt{2i\pi\hbar\tau/m}}\ ,
\end{eqnarray}
we can reduce in the one-dimension's $x$-direction the problem:
\begin{eqnarray}\label{Kx}
K_x(x,T+\tau;0,0)=\int_{b-a}^{b+a}dw\ \frac{e^{i\frac{m (x-w)^2}{2\hbar\tau}}}{\sqrt{2i\pi\hbar\tau/m}}
\frac{e^{i\frac{m w^2}{2\hbar T}}}{\sqrt{2i\pi\hbar T/m}}\ .
\end{eqnarray}
Then we can reduce the dimension of the problem keeping only the $x$-axis
propagator, as you can see in the References\cite{FH},\cite{Frabboni}.
In the Reference,\cite{Frabboni} they consider the slits in 
the $x$-$y$-plane, in two-dimensions and not in one-dimension 
as we do in the present article.\\

If we take the limit $a\rightarrow\infty$ (infinite slit), 
by (\ref{K2DFormula}) and by (\ref{GaussianConvo}) we get:
\begin{eqnarray*}
K_{\infty}^{(2D)}((x,L+D),T+\tau;(0,0),0)&=&\frac{e^{i\frac{mD^2}{2\hbar T}}}{\sqrt{2i\pi\hbar T/m}} 
\frac{e^{i\frac{m L^2}{2\hbar\tau}}}{\sqrt{2i\pi\hbar\tau/m}}
\int_{-\infty}^{\infty}dw\ \frac{e^{i\frac{m (x-w)^2}{2\hbar\tau}}}{\sqrt{2i\pi\hbar\tau/m}}
\frac{e^{i\frac{m w^2}{2\hbar T}}}{\sqrt{2i\pi\hbar T/m}}
{}\\{}&=&\frac{e^{i\frac{mD^2}{2\hbar T}}}{\sqrt{2i\pi\hbar T/m}} 
\frac{e^{i\frac{m L^2}{2\hbar\tau}}}{\sqrt{2i\pi\hbar\tau/m}}
\frac{e^{i\frac{m x^2}{2\hbar (T+\tau)}}}{\sqrt{2i\pi\hbar (T+\tau/m)}}\ .
\end{eqnarray*}
Then we find the amplitude corresponding to the transition
between the position $(x=0,z=0)$ at the time $t=0$ 
and the position $(x,z=L+D)$ at the time $t=T+\tau$, 
knowing that the position in the $z$-direction is $D$ at the time $t=T$ :
\begin{eqnarray*}
K_{\infty}^{(2D)}((x,L+D),T+\tau;&&(0,0),0)
{}\\{}&&=K_{0,x}\left(x,T+\tau;0,0\right)K_{0,z}\left(L+D,T+\tau;D,\tau\right)K_{0,z}\left(D,T;0,0\right)\ ,
\end{eqnarray*}
where $K_{o,X}(X',t';X,t)=(m/2i\pi\hbar(t'-t))^{1/2}\exp{\left(im(X'-X)^2/(2\hbar(t'-t))\right)},\ X=x,z,$ 
are the 1-dimensional free propagators.
Then, integrating over the intermediate positions $D$ between $-\infty$ and $+\infty$,
we find the two dimensional free propagator (\ref{K2D}) between $(x=0,z=0)$ at $t=0$
and $(x,z=L+D)$ at the time $t=\tau+T$:
\begin{eqnarray*}
\int_{-\infty}^{\infty}dD\ K_{\infty}^{(2D)}&&((x,z=L+D),T+\tau;(0,0),0)
{}\\{}&&=\int_{-\infty}^{\infty}dD\ \frac{e^{i\frac{mD^2}{2\hbar T}}}{\sqrt{2i\pi\hbar T/m}} 
\frac{e^{i\frac{m (z-D)^2}{2\hbar\tau}}}{\sqrt{2i\pi\hbar\tau/m}}
\frac{e^{i\frac{m x^2}{2\hbar (T+\tau)}}}{\sqrt{2i\pi\hbar (T+\tau)/m}}\ ,
{}\\{}&&=\left(\frac{m}{2i\pi\hbar (T+\tau)}\right)e^{i\frac{m (x^2+z^2)}{2\hbar (T+\tau)}}
{}\\{}&&= K_0^{(2D)}((x,z),T+\tau;(0,0),0)\ ,
\end{eqnarray*} 
to get the result, we have used the convolution formula between two Gaussians (\ref{GaussianConvo}).

\section{The exact result in terms of a Fresnel integral}

Now we want to compute the amplitudes $A_1(x;a,b),\ A_2(x;a,b)$
at each point $x$ on the screen, using the Feynman formulation,
and then to add both amplitudes to get the total amplitude
$A(x;a,b)$ and finally to take the square of the modulus,
obtaining the probability $P(x;a,b)\equiv |A(x;a,b)|^2$.

By (\ref{Kx}), the formal expression for $A_1(x;a,b)$ is :
\begin{eqnarray}\label{A1}
A_1(x;a,b)=\int_{-\infty}^{+\infty} dw\ \chi_{[b-a,b+a]}(w)
K_0(x,T+\tau;w,T) K_0(w,T;0,0)\ ,
\end{eqnarray}
where $T$ is the travel time of the electron from the source to
the slit and $\tau$ from the slit to the screen. 
Recall that to obtain formula
(\ref{A1}), we used a similar argument to that which enabled us to
write the formula (\ref{Klim}), writing the integral over all
possible paths as the product of independent amplitudes at
successive times. However, in this case, at time $ T $, we have to
integrate over a finite interval (the slit), which results in the
more complicated expression (\ref{A1}) rather than a Gaussian:

\begin{figure}
 \centering
\includegraphics[width=130mm]{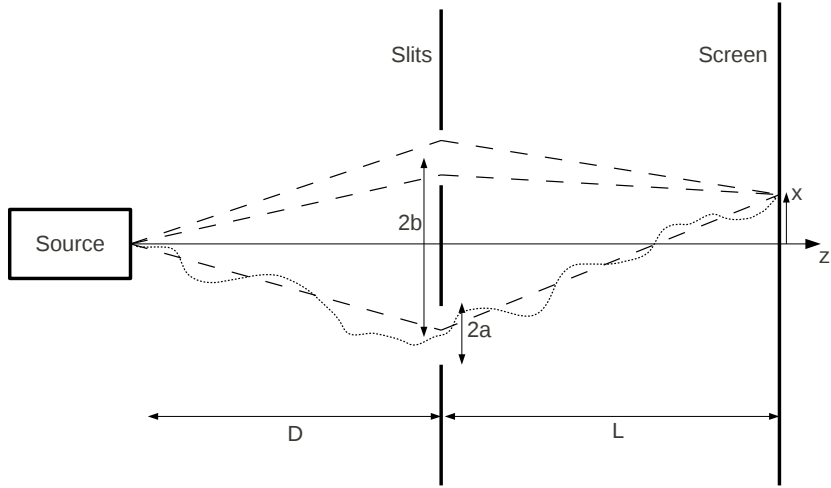}
\caption{\label{Fig.1} Schematic depiction of the apparatus with
the source, the two-slits and the screen.
We have represented four differents paths going through the slits, from the source to the screen.
{The dashed lines represent three classical paths and the thin dashed line a curved path.
The Feynman path Integral formulation allows one to compute the propagator 
summing over all different paths, see equations 
(\ref{Klim}), (\ref{Kfree}), (\ref{A1}), (\ref{A1C2}), (\ref{A2C2}), (\ref{A}).}}
\end{figure}

\begin{eqnarray}\label{A11}
A_1(x;a,b)=\int_{b-a}^{b+a} dw \frac{1}{\sqrt{2i\pi
\hbar\tau/m}}e^{i\frac{m (x-w)^2}{2\hbar\tau}}
\frac{1}{\sqrt{2i\pi \hbar T/m}}e^{i\frac{m w^2}{2\hbar T}}  \ .
\end{eqnarray}
Notice that :
\begin{eqnarray*}
\frac{m(x-w)^2}{2\hbar\tau}+\frac{mw^2}{2\hbar T}=
(\frac{m}{2\hbar\tau}+\frac{m}{2\hbar
T})(y-\frac{x}{1+\tau/T})^2 + \frac{mx^2}{2\hbar(T+\tau)}\ ,
\end{eqnarray*}
and hence,
\begin{eqnarray}\label{A1C1}
A_1(x;a,b)&=&\frac{e^{i\frac{mx^2}{2\hbar
(T+\tau)}}}{\sqrt{2i\pi \hbar (T+\tau)/m}}\int_{b-a}^{b+a} dw
\sqrt{\frac{T+\tau}{2i\pi \hbar
T\tau/m}}\exp{\left(i(\frac{T+\tau}{2\hbar
T\tau/m})(w-\frac{x}{1+\tau/T})^2\right)}  \ .
{}\nonumber\\{}&=&\frac{e^{i\frac{mx^2}{2\hbar
(T+\tau)}}}{\sqrt{(2i)^2\pi \hbar
(T+\tau)/m}}\int_{\alpha_{-}^{(1)}(x)}^{\alpha_{+}^{(1)}(x)} dw'
\exp{\left(\frac{i\pi}{2} w'^2\right)},
\end{eqnarray}
where
\begin{eqnarray}\label{alpha1}
&&\alpha_{\pm}(x;a,b)\equiv\sqrt{\frac{(T+\tau)}{\pi \hbar
T\tau/m}}(b\pm a)-\frac{x}{\sqrt{\pi
\hbar\tau/m}}\sqrt{\frac{T}{T+\tau}}.
\end{eqnarray}
In (\ref{A1C1}), we see that we have the integral of a Gaussian
with complex argument. Decomposing the integral in real and
imaginary parts, we get two integrals of cosine and sine functions
respectively, with second degree polynomial arguments. These
integrals are the well-known \textit{Fresnel functions}
\cite{Abramowitz}:
\begin{eqnarray*}
&& C[u]\equiv\int_0^u dw \cos{(\frac{\pi w^2}{2})}\ , {}\\{}&&
S[u]\equiv\int_0^u dw \sin{(\frac{\pi w^2}{2})}\ .
\end{eqnarray*}
Thus we obtain explicit analytical expressions for the amplitudes:
\begin{eqnarray}\label{A1C2}
 A_1(x;a,b)&&=\frac{e^{i\frac{m x^2}{2\hbar
(\tau+T)}}}{\sqrt{(2i)^2\pi \hbar (T+\tau)/m}} \times
{}\nonumber\\{}&&\big(C[\alpha_{+}(x;a,b)] - C[\alpha_{-}(x;a,b)] +
iS[\alpha_{+}(x;a,b)] - iS[\alpha_{-}(x;a,b)] \big)\ ,
\end{eqnarray}
\begin{eqnarray}\label{A2C2}
A_2(x;a,b)&=&A_1(x;a,-b)\ .
\end{eqnarray}
{For two slits, we can compute the total amplitude by summing the amplitudes for both slits:
\begin{eqnarray}\label{A}
A(x;a,b)=A_1(x;a,b)+A_2(x;a,b)\ .
\end{eqnarray}}

\section{Physical parameters, approximations and interpretations}

Recall that the length of the slits and the distance between them
is small compared to the horizontal distances $D$ and $L$, hence we can
assume that the wave length of the electron $\lambda=h/mv$ 
is approximatively given by $h/mv_z$, where $v_z=L/\tau=D/T$. 
We will use this expression for the wave length in the following.

\subsection{Diffraction by a single slit}

We can write for the single-slit case (of size $2a$), the
analogue of the function defined by (\ref{alpha1}):
\begin{equation}\label{alpha}
\alpha(x;a)=\sqrt{N_F(a)}\sqrt{1+L/D}
\left(1-\frac{x}{a}\frac{1}{1+L/D}\right)
\end{equation}
where $N_F(a)=2a^2/\lambda L$ is the \textit{Fresnel number}.

We can now easily compute the single-slit diffraction
probability:
\begin{eqnarray}\label{P1}
&&P^{(1Slit)}(x;a) = |A_1(x;a,b=0)|^2
{}\nonumber\\{}&=&\frac{1}{2\lambda(L+D)} \left([C(\alpha(x;a)) +
C(\alpha(x;-a))]^2 + [S(\alpha(x;a)) + S(\alpha(x;-a))]^2 \right)\ .
\end{eqnarray}

Let us introduce the following parameters $\eta\equiv 1+L/D$ and
$\gamma=\eta-1$. We can then plot the functions (\ref{P1}) for
various values of these parameters, see Fig 3. Note that the
Fresnel functions in (\ref{P1}) behave differently depending on
the value of the Fresnel number $N_F(a)$, esp. depending on
whether it is greater or less than unity. To understand these
differences explicitly, we will analyse the asymptotic behavior
of these functions for different regimes of $N_F(a)$ using the
known asymptotics of the Fresnel functions: (see \cite{Abramowitz})
\begin{eqnarray}\label{ApproxCS}
&& C(\pm u)\simeq\ \pm\frac{1}{2}+\frac{1}{\pi u}\sin{\frac{\pi
u^2}{2}}\ ,\ u\gg1\ , {}\nonumber\\{}&& S(\pm u)\simeq\
\pm\frac{1}{2}-\frac{1}{\pi u}\cos{\frac{\pi u^2}{2}}\ ,\ u\gg1\ .
\end{eqnarray}

{
If we have: $$\frac{x}{a\eta}-1\gg \frac{1}{\sqrt{N_F(a)\eta}}\ ,$$ 
we get the asymptotic behavior for the functions defined by (\ref{alpha}):
\begin{eqnarray}\label{alphasympt0}
\alpha(x;a)\ll-1\ \mathrm{and}\ \alpha(x;-a)\gg1 \Leftrightarrow \pm\alpha(x;\mp a)\gg1\ ,
\end{eqnarray}
and so by (\ref{ApproxCS}) and (\ref{alphasympt0}),
the asymptotics formulas of the Fresnel functions in (\ref{P1}) are given by:
\begin{eqnarray}\label{CSasympt1}
&& C[\alpha(x;\pm a)]\simeq \pm\frac{1}{2} + \frac{1}{\pi\alpha(x;\pm a)}\sin{(\frac{\pi\alpha(x;\pm a)^2}{2})},
{}\nonumber\\{}&& S[\alpha(x;\pm a)]\simeq \pm\frac{1}{2} - \frac{1}{\pi\alpha(x;\pm a)}\sin{(\frac{\pi\alpha(x;\pm a)^2}{2})}\ .
\end{eqnarray}}
{Then we get:
\begin{eqnarray}\label{FresnelFunctionApprox}
&& C[\alpha(x;+a)]+C[\alpha(x;-a)]\simeq \frac{1}{\pi\alpha(x;a)}\sin{(\frac{\pi\alpha(x;a)^2}{2})}
+\frac{1}{\pi\alpha(x;-a)}\sin{(\frac{\pi\alpha(x;-a)^2}{2})},
{}\nonumber\\{}&& S[\alpha(x;+a)]+S[\alpha(x;-a)]\simeq \frac{-1}{\pi\alpha(x;a)}\cos{(\frac{\pi\alpha(x;a)^2}{2})}
-\frac{1}{\pi\alpha(x;-a)}\cos{(\frac{\pi\alpha(x;-a)^2}{2})}
\end{eqnarray}}

Applying the Fresnel function asymptotic forms (\ref{FresnelFunctionApprox}) to
(\ref{P1}) and using the definition (\ref{alpha}), we deduce than if $N_F(a)\ll 1$ and
if $(x-a\eta)/a\eta\gg 1/\sqrt{N_F(a)\eta}\Leftrightarrow
x-a\eta\gg\sqrt{\lambda L/2}$, we get the following asymptotic
formula:
\begin{eqnarray}\label{P1ApproxNFsmall0}
P^{(1 Slit)}(x;a)&\simeq&\frac{2\gamma}{\pi^2\eta^2}
\left(\frac{a^2}{(\frac{x^2}{\eta^2}-a^2)^2} +
\frac{1}{\frac{x^2}{\eta^2}-a^2}\sin^2{(\pi N_F(a)
\frac{x}{a})}\right) \ ,
\end{eqnarray}

Moreover, since $N_F(a)\ll1$ and so $\frac{x}{a\eta}\gg1$, we have another asymptotic form
(large distance on the screen):
\begin{eqnarray}\label{P1ApproxNFsmall}
P^{(1 Slit)}(x;a)&\simeq&\frac{2\gamma}{\pi^2 x^2} \sin^2{(\pi
N_F(a) \frac{x}{a})},\ (N_F(a)\ll1,\ x/a\eta\gg 1)\ ,
\end{eqnarray}
In this case we are in the so-called \textit{Fraunhofer regime}
analogous to plane wave diffraction in optics.\cite{Optics} In
fact, notice that the distance between fringes is
$a/N_F(a)=\lambda_z L/2a$, c.f. Fig 3a.

Both approximations (\ref{P1ApproxNFsmall0}) and
(\ref{P1ApproxNFsmall}) are also valid if $N_F(a)$ is of the order
of unity (this is the \textit{intermediate regime}) provided that
$x\gg a\eta$. This means that the pattern on the screen far from
the position of the first lobe is well approximated by  equations
(\ref{P1ApproxNFsmall0}) and (\ref{P1ApproxNFsmall}), see Fig. 3b.

On the contrary, if $N_F(a)\gg 1$, we get different asymptotics
given by:
\begin{eqnarray}\label{P1ApproxNFhigh}
&&P^{(1
Slit)}(x)\simeq\frac{\gamma}{\eta}\left(\frac{\sqrt{N_F(a)}}{2a}+
\frac{\sin{(\frac{\pi}{2}N_F(a)\eta(1-\frac{x}{a\eta})^2)}}{2\pi\sqrt{\eta}(a-\frac{x}{\eta})}
+\frac{\sin{(\frac{\pi}{2}N_F(a)\eta(1+\frac{x}{a\eta})^2)}}{2\pi\sqrt{\eta}(a+\frac{x}{\eta})}\right)^2
{}\nonumber\\{}&+&\frac{\gamma}{\eta}\left(\frac{\sqrt{N_F(a)}}{2a}-
\frac{\cos{(\frac{\pi}{2}N_F(a)\eta(1-\frac{x}{a\eta})^2)}}{2\pi\sqrt{\eta}(a-\frac{x}{\eta})}
-\frac{\cos{(\frac{\pi}{2}N_F(a)\eta(1+\frac{x}{a\eta})^2)}}{2\pi\sqrt{\eta}(a+\frac{x}{\eta})}\right)^2,\ |x|<a\eta,
\end{eqnarray}
\begin{eqnarray}\label{P1ApproxNFhigh2}
P^{(1 Slit)}(x)&&
\simeq\frac{2\gamma}{\pi^2\eta^2}\left(\frac{a^2}{(\frac{x^2}{\eta^2}-a^2)^2}+\frac{1}{\frac{x^2}{\eta^2}-a^2}\sin^2{(\pi
N_F(a) \frac{x}{a})}\right) ,\ |x|>a\eta\ ,
\end{eqnarray}
since the asynptotic behavior of the functions (\ref{alpha}) are:
\begin{eqnarray}
&&\alpha(x;\pm a)\gg1,\ \mathrm{if}\ N_F(a)\gg1\ \mathrm{and}\ |x|<a\eta\ ,\label{alphasympt1}
{}\\{}&& \pm\alpha(x;\mp a)\gg1,\ \mathrm{if}\ N_F(a)\gg1\ \mathrm{and}\ |x|>a\eta\ ,\label{alphasympt2}
\end{eqnarray}
so if $|x|<a\eta$, by (\ref{alphasympt1}) and (\ref{ApproxCS}), we get:
\begin{eqnarray}\label{CSasympt2}
&& C[\alpha(x;\pm a)]\simeq \frac{1}{2} + \frac{1}{\pi\alpha(x;\pm a)}\sin{(\frac{\pi\alpha(x;\pm a)^2}{2})},
{}\nonumber\\{}&& S[\alpha(x;\pm a)]\simeq \frac{1}{2} - \frac{1}{\pi\alpha(x;\pm a)}\sin{(\frac{\pi\alpha(x;\pm a)^2}{2})}\ ,
\end{eqnarray}
then by (\ref{alphasympt1}) and (\ref{CSasympt2}), we have:
\begin{eqnarray*}
&& C[\alpha(x;+a)]+C[\alpha(x;-a)]\simeq 1+\frac{1}{\pi\alpha(x;a)}\sin{(\frac{\pi\alpha(x;a)^2}{2})}
+\frac{1}{\pi\alpha(x;-a)}\sin{(\frac{\pi\alpha(x;-a)^2}{2})},
{}\nonumber\\{}&& S[\alpha(x;+a)]+S[\alpha(x;-a)]\simeq 1-\frac{1}{\pi\alpha(x;a)}\cos{(\frac{\pi\alpha(x;a)^2}{2})}
-\frac{1}{\pi\alpha(x;-a)}\cos{(\frac{\pi\alpha(x;-a)^2}{2})}\ ,
\end{eqnarray*}
then we get (\ref{P1ApproxNFhigh}).

If $|x|>a\eta$, by (\ref{alphasympt2}) and (\ref{ApproxCS}), 
we get the same approximations as (\ref{CSasympt1}) and (\ref{FresnelFunctionApprox}),
then we obtain (\ref{P1ApproxNFhigh2}).

Note that the function (\ref{P1ApproxNFhigh}) oscillates rapidly
in the interval $[-a,+a]$ (esp. near the edges) around a constant
value $N_F(a)\gamma/2a^2\eta = 1/(\lambda_z (L+D))$, whereas for
$|x|>a\eta$, the function (\ref{P1ApproxNFhigh2}) decreases
rapidly to $0$. Hence $P^{(1 Slit)}$ tends at large $N_F(a)$ to
the \lq door\rq\ function defined by (\ref{Porte}), as might have
been expected, see Fig. 3c.

\begin{figure}[h]
\begin{center}
\scalebox{1}{\includegraphics{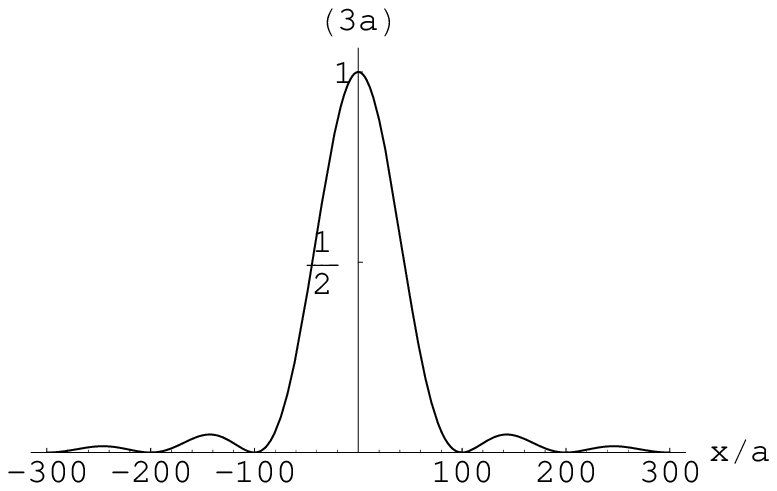}}\scalebox{1}{\includegraphics{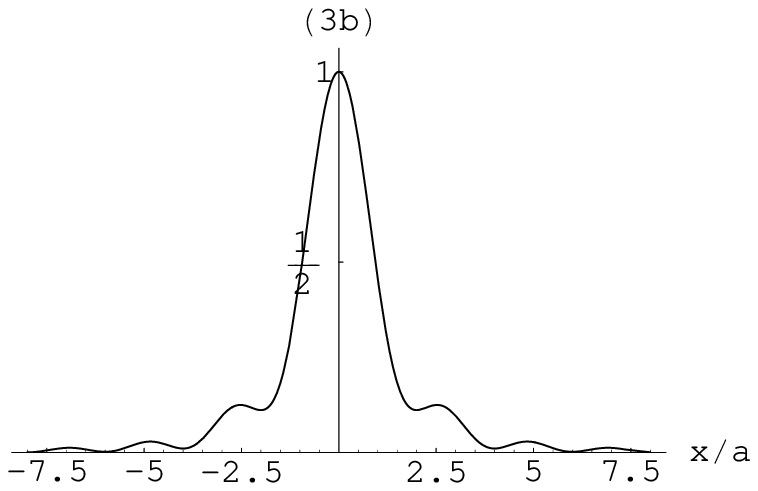}}
\scalebox{1}{\includegraphics{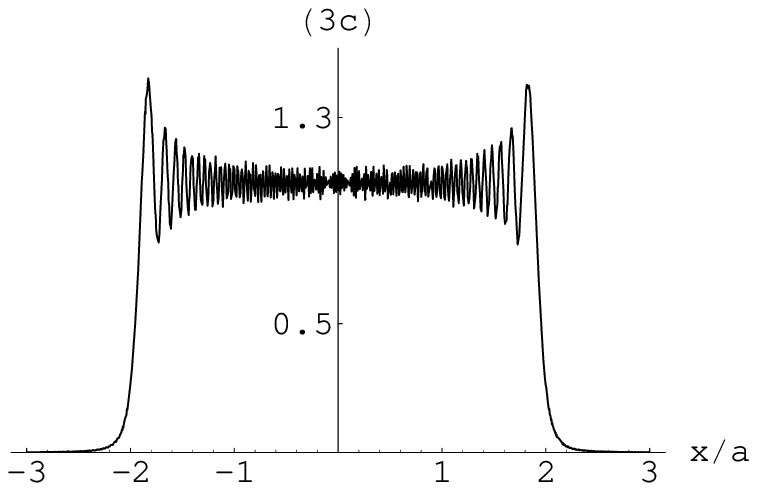}} \caption{\label{Fig.2}
Diffraction curve for a single-slit, with $\eta=2$. The
abscissae are the distances in units of $a$ and the ordinates
are the relative populations. We have $N_F(a)=0.01$ for the
Figure (3a), $0.5$ for (3b), and $100$ for (3c).}
\end{center}
\end{figure}

\subsection{Comment about the probability interpretation}

We can see that (\ref{P1}) has the physical dimension of the
inverse of a length squared and so it is neither a probability nor
a probability density. This apparent problem can, however, be seen
to be a matter of interpretation by looking at the formula
(\ref{ShroInt}). Indeed, the probability density for the
diffraction problem is given by $|\Psi(x,T+\tau)|^2$, where
$\Psi(x,T+\tau)$ is a \textit{normalized} wave function, i.e.
$\int_{R^1} dx\, |\Psi(x,T+\tau)|^2=1$. Therefore, we must choose
an initial wave function (at time $t=0$) which is also normalized
so that its square modulus describes the probability distribution
of the electron in the $x_0$ plane. Essentially, what we have done
in (\ref{P1}) is to take the initial wave function to be a
delta-function, whereas it should be the initial probability
distribution which is a delta-function, i.e. $\Psi(x,0)$ should be
\lq the square root of a delta-function'.

To make this clearer, consider for example a wave function at time
$t=0$ given by the square root of a Gaussian $\phi_\sigma(x_0) =
\frac{1}{(2\pi\sigma^2)^{1/4}} e^{-\frac{x_0^2}{4\sigma^2}}$, so
that $\int_{R^1} dx_0\,|\phi_\sigma(x_0)|^2 = 1$ and
$|\phi_\sigma(x_0)|^2 \rightarrow \delta(x_0)$ as $\sigma
\rightarrow 0$. In this way the wave function obtained in
(\ref{wavefunction}) is properly normalized so as to get the
probability of the presence of the electron at the point $x$ on
the screen by taking the square of the modulus. Indeed, the wave
function at time $t=T$, i.e. at the position of the slits, is
given by:
\begin{eqnarray}
\phi_\sigma(x,T) = \int_{R^1} dx_0\, K_0(x,T;x_0,0)
\phi_\sigma(x_0)\ ,
\end{eqnarray}
where  $K_0(x,T+\tau;x_0,0)$ is the free propagator defined by the
equation (\ref{Kfree}). Using the identity \begin{equation}
\label{K0deltarel}  \int_{R^1} K_0(x,T;x_0,0)^* K_0(x,T; x_0',0)
dx = \delta(x_0-x_0')
\end{equation} we have that $\phi_\sigma(x,T)$ remains normalized, i.e.
$\int |\phi_\sigma(x,T)|^2 dx = 1$.

The wave function at time $t=T+\tau$ is given by:
\begin{equation}
\Psi_\sigma(x,T+\tau) = \int_{-a}^{+a}dy
K_0(x,T+\tau;y,T)\phi_\sigma(y,T). \label{wavefunction}
\end{equation}
Now, the quantity of interest is the \textit{conditional}
probability (density) for the electron to be at the point $x$ on
the screen at the time $T+\tau$ \textit{given} that it was in the
interval $[-a,+a]$ at time $T$, i.e. given that it passed through
the slit:
\begin{eqnarray}\label{P}
P_\sigma\left(x,T+\tau \,|\,y\in [-a,+a],T\right) \equiv
\frac{|\Psi_\sigma(x,T+\tau)|^2}{\int_{-a}^{+a} dy\,
|\phi_\sigma(y,T)|^2}\ .
\end{eqnarray}
after which we wish to take the limit $\sigma\rightarrow0$. Using
the relation (\ref{K0deltarel}) one can see that the condition
probability (\ref{P}) is normalized so that this procedure just
amounts to division by a normalization factor. Thus
\begin{equation}\label{Z}
\int_{R^1} dx\, |\Psi_\sigma(x,T+\tau)|^2 = \int_{-a}^{+a}dy
|\phi_\sigma(y,T)|^2\ .
\end{equation}

To take the limit $\sigma \to 0$, note that $\phi_\sigma(x_0) =
(8\pi\sigma^2)^{1/4}g_{\sigma\sqrt{2}}(x_0)$ where
$g_{\sigma\sqrt{2}}(x_0)=\frac{e^{-\frac{x_0^{'2}}{4\sigma^2}}}{\sqrt{4\pi\sigma^2}}$
is a normalized Gaussian of variance $\sigma\sqrt{2}$, and hence
$g_{\sigma\sqrt{2}}(x_0) \rightarrow \delta(x_0),\
\sigma\rightarrow0$. Multiplying top and bottom of
(\ref{wavefunction}) by $(8\pi \sigma^2)^{1/2}$ we thus have
\begin{eqnarray}
&\lim_{\sigma\rightarrow0}&P_\sigma\left(x\in A,T+\tau|y\in
[-a,+a],T\right)= {}\nonumber\\{}&&=\lim_{\sigma\rightarrow 0}
\frac{(8\pi\sigma^2)^{-1/2}
|\Psi_\sigma(x,T+\tau)|^2}{(8\pi\sigma^2)^{-1/2}\int_{-a}^{+a}dy
|\phi_\sigma(y,T)|^2}\ , {}\nonumber\\{}&&= \frac{\left| \int dx_0
K(x,T+\tau; x_0,0) g_{\sigma \sqrt{2}}(x_0) \right|^2}{\int_{-a}^a
dy\, \left| \int dx_0 K_0(y,T;x_0,0) g_{\sigma \sqrt{2}} (x_0)
\right|^2} {}\nonumber\\{}&&= \frac{
|K(x,T+\tau;0,0)|^2}{\int_{-a}^{+a} dy |K_0(y,T;0,0)|^2}
{}\nonumber\\{}&&=\frac{\lambda L}{2a}  P^{1 Slit}(x;a)\ ,
\end{eqnarray}
where \begin{equation} \label{propagator} K(x,T+\tau;0,0) =
\int_{-a}^{+a} dy\, K_0(x,T+\tau;y,T) K_0(y,T;0,0) \end{equation}
is the propagator through the slit and $P^{(1 Slit)}(x;a) =
|K(x,T+\tau;0,0)|^2$ is given by (\ref{P1}). Note that this now
has the correct dimension of an inverse length.

\subsection{Interference and diffraction for two slits}
Similarly, one can find the two-slit diffraction probability
formula using (\ref{A1C2}), (\ref{A2C2}) :
\begin{eqnarray}\label{P2}
P^{(2 Slit)}(x;a,b) &=& P_1(x;a,b)+P_2(x;a,b)+I_{12}(x;a,b)\ ,
\end{eqnarray}
with the diffraction terms :
\begin{eqnarray}\label{P1P2}
&&P_1(x;a,b)= |A_1(x;a,b)|^2
{}\nonumber\\{}&=&\frac{\gamma}{2\lambda
L\eta}\left([C(\alpha_{+}(x;a,b))-C(\alpha_{-}(x;a,b))]^2+[S(\alpha_{+}(x;a,b))-S(\alpha_{-}(x;a,b))]^2\right)\
, {}\nonumber\\{}&& P_2(x;a,b) = |A_2(x;a,b)|^2 = P_1(x;a,-b)\ ,
\end{eqnarray}
and the interference term :
\begin{eqnarray}\label{I12}
I_{12}(x;a,b)&=&A_1(x;a,b)A_2(x;a,b)^* + A_2(x;a,b) A_1(x;a,b)^*
{}\nonumber\\{}&=&\frac{\gamma}{\lambda L\eta}
([C(\alpha_{+}(x;a,b))-C(\alpha_{-}(x;a,b))][C(\alpha_{+}(x;a,-b))-C(\alpha_{-}(x;a,-b))]
{}\nonumber\\{}&+&[S(\alpha_{+}(x;a,b))-S(\alpha_{-}(x;a,b))][S(\alpha_{+}(x;a,-b))-S(\alpha_{-}(x;a,-b))])\ .
\end{eqnarray}
Notice that there is an additional term compared to the
single-slit case called the \textit{interference term}, which is
of course quite similar to that in optics.\cite{Optics} This
results in a modulation effect of the curve given by (\ref{P2}) by
the sum of the diffraction terms (\ref{P1P2}) (modulo a
multiplicative factor), see Fig. 4.

Let us define the Fresnel numbers
$$ N_F(a)\equiv2a^2/\lambda_z L, \quad
N_F(b)\equiv2b^2/\lambda_z L \mbox{ and } N_F\equiv2ab/\lambda_z L =
\sqrt{N_F(a)N_F(b)/2}.  $$

Assume that the distance between the slits is large compared to
the size of the slits $b\gg a$. In the experiment considered one
fixes both parameters $a$ and $b$ and varies the distance between
the screen and the slits (keeping the same value for $\eta$).
Notice that because $b\gg a$, $N_F\gg 1$ does not necessarily
imply that $N_F(a)\gg1$. Thus we will see that both parameters
play different roles.

First, we establish the asymptotics of (\ref{P2}) for different
asymptotic value of $N_F(a)$. Under the condition $N_F(a)\ll 1$
and at large scales $|x-b\eta|\gg a\eta$ et $|x+b\eta|\gg
a\eta$, we get  similar expressions for $P_1(x;a,b)$ and
$P_2(x;a,b)$ as (\ref{P1ApproxNFsmall}). We have to compute the
asymptotic expression for the interference term. This yields
\begin{eqnarray}\label{P2slApprox}
P^{(2 Slit)}(x;a,b)&\simeq&\frac{2\gamma}{\pi^2 (x-b\eta)^2}
\sin^2{\left(\pi N_F\eta(1-\frac{x}{b\eta})\right)}
+\frac{2\gamma}{\pi^2 (x+b\eta)^2} \sin^2{\left(\pi
N_F\eta(1+\frac{x}{b\eta})\right)}
{}\nonumber\\{}&-&\frac{\gamma}{\pi^2 (x^2-b^2\eta^2)}
\left(\cos{(2\pi (N_F+N_F(a))\frac{x}{a})}-\cos{(2\pi
N_F\eta(1+\frac{x}{a\eta}))}\right)
{}\nonumber\\{}&-&\frac{\gamma}{\pi^2 (x^2-b^2\eta^2)}
\left(\cos{(2\pi (N_F-N_F(a))\frac{x}{a})}-\cos{(2\pi
N_F\eta(1-\frac{x}{a\eta}))}) \right)\ .
\end{eqnarray}

%{
One can observe that there are two phases: the
separated phase ($N_F\gg 1$) and the mixed phase ($N_F\ll 1$). At
the same time there are the Fresnel and Fraunhofer regimes
depending on the values of $N_F(a)$ as explained above. The
distinction between two phases is purely geometric and
characterizes the separation respectively mixture of diffraction
curves. Indeed, similar to optics, to observe the two diffraction
curves separately, the fringe modulation $\lambda_z L/2a$ must be
less than the distance between the origins of the two curves
(being centered in $\pm b\eta$) because otherwise both curves are
mixed. Thus the criterion is written $ \lambda_z L/2a < b\eta $
and therefore $N_F\eta > 1$. This obviously unlike the case $ N_F
\ll 1 $ where the two curves are combined (one added to the other)
and where we observe modulation interference. %}

%{
To see this more formally, consider firstly the case
$N_F\eta\ll1$. If $|x|>\lambda L/2a$ then $|x|\gg b\eta$ and we
can give an approximation of (\ref{P2slApprox}). Indeed, the first
two terms are approximately equal and contribute $$
\frac{4\gamma}{\pi^2 x^2} \sin^2 ( \frac{2 \pi a}{\lambda L} x).
$$  In the last two terms we develop the cosine functions and get
\begin{eqnarray*}
&& 2 \left[\cos{(\frac{2\pi b}{\lambda L}x)}\cos{(\frac{2\pi
a}{\lambda L}x)}-\cos{(\frac{2\pi ab\eta}{\lambda
L})}\cos{(\frac{2\pi b}{\lambda L}x)} \right] \\ && \approx -4
\cos (\frac{2 \pi b}{\lambda L} x) \sin^2 (\frac{2 \pi a}{\lambda
L} x),
\end{eqnarray*}
where we used $\cos{(\frac{2\pi ab\eta}{\lambda L})}\simeq1$
because $N_F\eta\ll1$. Adding the terms we obtain
\begin{eqnarray}\label{P2slApprox2}
P^{(2slit)}(x;a,b) \simeq \frac{8\gamma}{\pi^2
x^2}\sin^2{(\frac{2\pi a}{\lambda L}x)} \cos^2{(\frac{2\pi
b}{\lambda L}x)}\ .
\end{eqnarray}
This is the familiar optical formula: see Reference,\cite{Optics} formula
(10) Chap. VIII-6. It shows that the diffraction curves are modulated by
interference fringes. The distance between two interference
fringes is of the order of $\lambda_z L/2b$ whereas that between
minima of the diffraction curves is of the order of $\lambda_z
L/2a \gg \lambda_z L/2b$, see Fig. 4a, Fig. 4b (far from the first
lobe) and Fig. 4c.

%{
Secondly, consider the case that $N_F\eta\gg1$ (while still
assuming $N_F(a)\ll 1$). If $x-b\eta>-\lambda L/2a$ (or
$x+b\eta<\lambda L/2a$) then $(x-b\eta)^{-2}\gg (x+b\eta)^{-2}$
(respectively $(x-b\eta)^{-2}\ll (x+b\eta)^{-2}$) so one of the
two terms is negligible in the respective domain. Moreover, in
both cases, the interference term is small compared to the
diffraction term since $(x-b\eta)^{-2}\gg (x^2-b^2\eta^2)^{-1}$.
The total probability is therefore approximatively equal to a sum
of the two diffraction curves centered at $\pm b\eta$ modulated by
an interference term which oscillates rapidly with a relatively
small amplitude, c.f. Fig 4c :
\begin{eqnarray} \label{P2slApprox3}
P^{(2 slit)}(x;a,b) &\simeq & \frac{2\gamma}{\pi^2 (x\mp b\eta)^2}
\sin^2 \left(\pi N_F \eta (1\mp\frac{x}{b\eta}) \right) +
O(\frac{1}{x^2-b^2\eta^2}),\ |x\pm b\eta| > \frac{\lambda L}{2a}\
.
\end{eqnarray}
%}

Now, consider $N_F(a)\gg 1$. Since, unlike the previous cases, we
do not need special conditions for the position $x$ on the screen,
we find similar formulas to (\ref{P1ApproxNFhigh}) and
(\ref{P1ApproxNFhigh2}) for the direct terms $P_1$ and $P_2$,
except that $x$ is replaced by $x-b\eta$ for the slit centered at
$+b$ and by $x+b\eta$ for the slit centered at $-b$.

For the interference term, inserting the asymptotics (\ref{ApproxCS})
into (\ref{I12}) above results in the sum of two terms, one being
the product of differences of cosin-functions, the other the product
of differences of sin-functions. The problem is obviously
symmetric about $x=0$, so we need only
consider the case $x>0$. Then there are again two cases: \\
(i)\ $|x-b| > a$; and (ii)\ $|x-b| < a$. \\
In the first case, both terms decrease like $1/(x+b)(x-b)$ with
various fluctuating factors as in (\ref{P2slApprox}). In the
second case, $P_1$ behaves as in (\ref{P1ApproxNFhigh}) but
centred around $x=b$, and the other terms are negligible. We do
not write the asymptotic formulas explicitly because the result is
simply the observation that in this case we obtain a sum of two
separated diffraction curves in the Fresnel regimes,
i.e. curves that tend to the door functions in the limit, see Fig. 4d.\\

Notice an interesting behavior of the interference pattern in Fig.
4c, where we see that the interference amplitudes are very small
compared to the diffraction amplitude inside a band $25 {< \atop
\sim} |x/a| {< \atop \sim} 75$, so that there are no interference
fringes. This is also discernible in Fig. 4b of the Reference\cite{Frabboni}
which corresponds to the calculated two-slit diffraction images,
where one can observe the absence of fringes in a band. However,
this phenomenon is not apparent on the corresponding experimental
image. This is probably due to the difference in defocussing
between the calculated and experimental images, see Fig. 3b and
Fig. 4b of the Reference.\cite{Frabboni} Indeed, the existence of such a band
is quite sensitive to the value of the parameter $N_F(a)$.

\begin{figure}[h]
\begin{center}
\scalebox{1}{\includegraphics{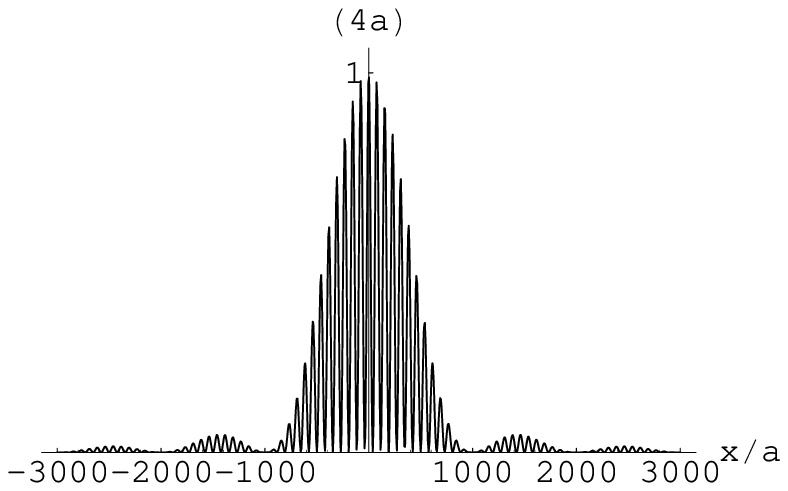}}\scalebox{1}{\includegraphics{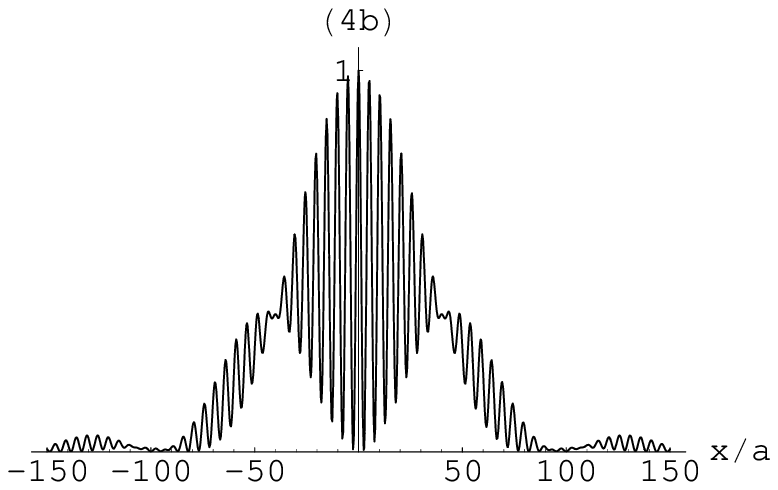}}
\scalebox{1}{\includegraphics{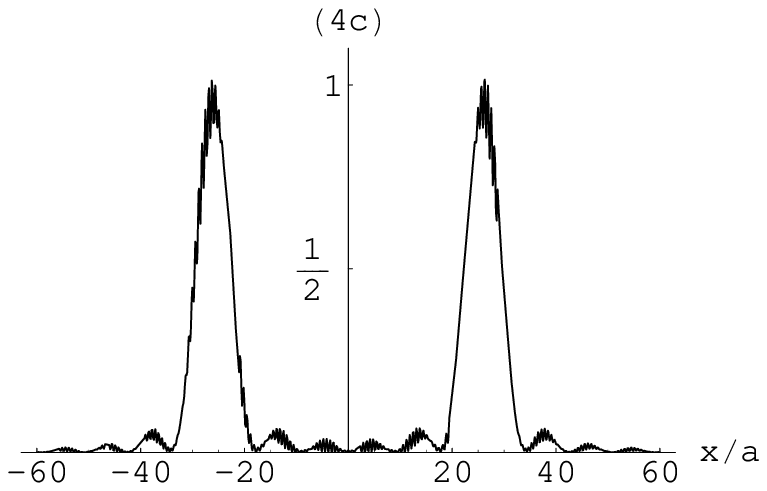}}\scalebox{1}{\includegraphics{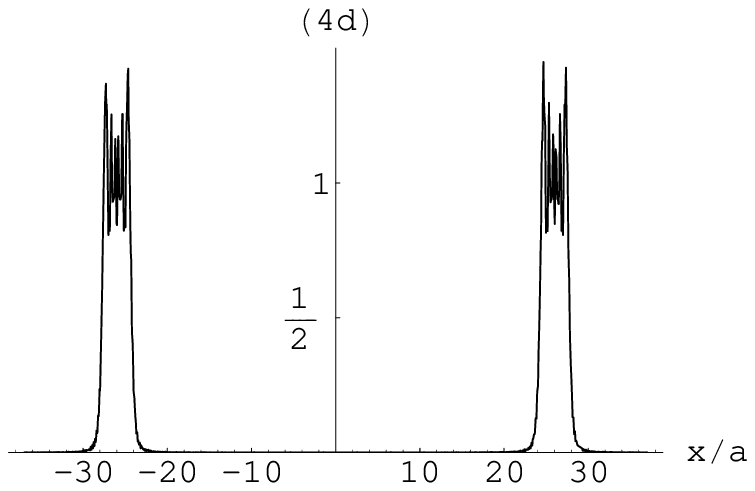}}
\caption{\label{Fig.3} Interference curves $(\ref{P2})$ from two
slits, with $b/a=13,\ \eta=2$. The abscissae are the distances in
units of $a$ and the ordinates are the relative populations. We
have $N_F(a)=0.001$ for the Figure (4a) , $0.015$ for (4b), $0.12$
for (4c) and $6$ for (4d).}
\end{center}
\end{figure}

\section{Conclusion and remarks}

- We have briefly presented the Feynman approach to quantum
mechanics, based on the Lagrangian formulation of classical
mechanics, and the associated change in paradigm in the transition
from classical to quantum mechanics. We note that in this
approach, the transition from classical to quantum mechanics is
quite natural because it relies mostly on concepts well known to
students of analytical mechanics and does not confuse particle and
wave behavior. It thus avoids some metaphysical questions and
leads directly to the solution of the diffraction and interference
problems above, and hence to a better understanding of the quantum
mechanics of such quintessential phenomena. This justifies
introducing the Feynman formulation at an early stage especially
as the semi-classical approach that is often used in a first
course relies on the idea of  quantification of the action. A
parallel introduction of the Feynman integral thus makes sense as
it clarifies the passage classical to quantum.

 - Secondly, it seemed of interest to derive explicit formulas for the problem
of diffraction / interference by one or two slits, and to discuss
the results based on the physical parameters of the system,
notably the Fresnel numbers and the distance scale at which we
observe on the screen. The properties of the diffraction and
interference patterns are not apparent from the exact formulas
(\ref{P1}) (\ref{P2}), so it is useful to establish asymptotic
forms (\ref{P1ApproxNFsmall0}) (\ref{P1ApproxNFsmall})
(\ref{P1ApproxNFhigh}) for the case of
one slit, and (\ref{P2slApprox}) for the case of two slits.\\

We summarize the various conclusions. In case of a single slit:

 - If $N_F(a)\ll 1$, this is the \textit{Fraunhofer regime} for which the distribution
curve is similar to the plane wave case, \textit{c.f.} equations
(\ref{P1ApproxNFsmall0}), (\ref{P1ApproxNFsmall}) and Fig. 3a.

 - If $N_F(a)\gg1$, this is the \textit{Fresnel regime} for which the diffraction
curve approximates the form of the slit, c.f. equations
(\ref{P1ApproxNFhigh}), (\ref{P1ApproxNFhigh2}) and Fig. 3c.

 - If $N_F(a)\sim 1$ one is in the \textit{intermediate regime} for which there is
a spreading around the center of the electronic distribution and
we find the case of Fraunhofer distances on the
screen, c.f. equation (\ref{P1}) and Fig. 3b. \\

In the case of two slits of width $ 2a $, and separated by a
distance $2b$ with $ b\gg a $, we can make similar distinctions as
in the one-slit diffraction case but there is also a transition
between two phases dependent on the optical resolution:

- If $N_F\ll 1$, one is in the \textit{mixed
phase}, i.e. we observe an interference curve modulated by a
diffraction curve for a slit of size $ a $ in this case $ N_F(a)
\ll 1$, then we are in the regime of Fresnel, c.f. equations
(\ref{P2slApprox}), (\ref{P2slApprox2}) and Fig. 4a.

- If $N_F\gg 1$, one is in the
\textit{separated phase}, and there are two interference curves
(the interference amplitudes are lower) modulated by the
diffraction curves corresponding to both slits, each curve being
centered respectively at $ \pm b\eta $; the shapes of the
diffraction curves modulating the signals of each of the slits
depends on $ N_F(a) $ and are similar to the case of a single slit
as summarized above (with three regimes: Fresnel, Fraunhofer and
intermediate), c.f. (\ref{P2slApprox3}), Fig. 4c for $N_F(a)\ll1$ 
and Fig. 4d for $N_F(a)\gg1$.

 - If $N_F \sim 1$ we observe a separation between two interference curves,
modulated by the diffraction curve corresponding to one slit at the intermediate regime.
see Fig. 4b and Eq. (\ref{P2slApprox}).  

Note that the fringes corresponding to the diffraction are at a
distance $\lambda L/2a$ and those for interference at about
$\lambda L/2b$. The analytical properties of our asymptotics of
two slits do not permit us to estimate these distances more
exactly, but
by analogy with optics they may be considered adequate. \\

In perspective, I suggest to take into account the quantum-mechanical way 
in the $z$-direction to solve the problem completely.
Indeed, recall that as we discussed in the Section III,
we consider in this article that the problem is separated
in two motions, one between the source and the slits 
and the other one between the slits and the screen,
which is rigorously not true. 
This is a challenging task
since we have also to compute the loop path corrections,\cite{Yabuki} 
but it could be an interesting contribution to the European Journal of Physics.

\section*{acknowledgments}
I would like to thank the Professor Tony Dorlas 
for discussions, encouragements and English corrections of the manuscript.

\end{document}